\documentclass[procedia]{easychair}

\usepackage{doc}
\usepackage{makeidx}

\usepackage{times}
\usepackage{graphicx}
\usepackage{epstopdf}
\usepackage{booktabs} 
\usepackage{subfigure}

\usepackage{listings} 
\usepackage{xcolor} 
\def\procediaConference{99th Conference on Topics of
  Superb Significance (COOL 2014)}

\title{On the Accelerating of Two-dimensional Smart Laplacian Smoothing on the GPU}

\titlerunning{Accelerating Smart Laplacian Smoothing on the GPU \ldots}

\author{
    Kunyang Zhao\inst{1}
\and
    Gang Mei\inst{1,2}\thanks{The author declares that this paper has been submitted to the International Conference on Computational Science ICCS 2015. Corresponding author: \email{gangmeiphd@gmail.com}}
\and
    Nengxiong Xu\inst{1}
\and
    Jiayin Zhang\inst{1}\\
}

\institute{
  School of Engineering and Technology, China University of Geosciences, 100083, Beijing, China\\
  \email{\{kunyang,xunengxiong,zhangjy\}@cugb.edu.cn}
\and
   Institute of Earth and Environmental Science, University of Freiburg, Albertstr.23B, D-79104, Freiburg im Breisgau, Germany\\
 }

\authorrunning{K. Zhao, G. Mei, N.Xu and J. Zhang}

\begin{document}

\maketitle


\begin{abstract}
    This paper presents a GPU-accelerated implementation of 
  	two-dimensional Smart Laplacian smoothing. This implementation is developed 
  	under the guideline of our paradigm for accelerating Laplacian-based mesh 
  	smoothing \cite{1mei2014}. Two types of commonly used data layouts, 
  	Array-of-Structures (AoS) and Structure-of-Arrays (SoA) are used to 
  	represent triangular meshes in our implementation. Two iteration forms that 
  	have different choices of the swapping of intermediate data are also 
  	adopted. Furthermore, the feature CUDA Dynamic Parallelism (CDP) is employed 
  	to realize the nested parallelization in Smart Laplacian smoothing. 
  	Experimental results demonstrate that: (1) our implementation can achieve 
  	the speedups of up to 44x on the GPU GT640; 
  	(2) the data layout AoS can always obtain better efficiency than the SoA 
  	layout; (3) the form that needs to swap intermediate nodal coordinates is 
  	always slower than the one that does not swap data; (4) the version of our 
  	implementation with the use of the feature CDP is slightly faster than the 
  	version where the CDP is not adopted.
\end{abstract}


%
%

\section{Introduction}
\label{sec:introduction}

The generation of computational mesh models plays a key role in numerical simulation. The 
quality of meshes strongly affects the computational efficiency and 
the accuracy of numerical results. In general, mesh models are needed to be 
optimized to improve the mesh quality after initially creating. There are 
typically two categories of mesh optimization methods: (1) the first one is 
mesh clear-up / modification, which changes the topology of meshes to 
improve the mesh quality \cite{2zegard,3damato,4chen}; and (2) the other one is the mesh smoothing, 
which leaves the mesh connectivity unchanged but to relocate the position of 
mesh nodes to enhance the mesh quality \cite{5damato2013,6dahal,7vartz}. A number of mesh smoothing 
methods have been introduced and widely used in various applications. Among 
the mesh smoothing methods, the Laplacian-based and the optimization-based 
methods are the two types of the most frequently used mesh smoothing 
approaches in practice \cite{9canann,10vartz}.

The basic idea behind the Laplacian mesh smoothing is quite simple. In each 
iteration, the new nodal position of each smoothed points is directly the 
geometric center of those of its adjacent / neighboring nodes \cite{11field1988}. 
There are no other complex smoothing operations. Thus, the Laplacian mesh 
smoothing is computationally non-intensive, and frequently used in various 
applications. 

However, when smoothing large meshes consisting of a large number of nodes 
and elements, the computational cost is still too high. To improve the efficiency, an effective strategy is 
to perform the mesh smoothing in parallel. For example, Mei, et al. \cite{1mei2014} developed a generic paradigm for accelerating the 
Laplacian-based mesh smoothing on the GPU. Dahal and Newman \cite{6dahal}
described efficient GPU-accelerated implementations of three triangular 2D 
mesh smoothing algorithms. In addition, D'Amato and Venere \cite{5damato2013} presented an implementation of a non-Laplacian smoothing method on the 
GPU to optimize tetrahedral meshes in parallel. Other recent efforts 
attempting at parallelizing mesh smoothing include those work presented in \cite{13beniterz,14benitez,Gorman2012}.

In this paper, we presents an efficient GPU-accelerated implementation of 
two-dimensional Smart Laplacian smoothing. The presented implementation is 
developed under the guideline of our previously proposed paradigm for 
accelerating Laplacian-based mesh smoothing \cite{1mei2014}. Two types of 
commonly used data layouts, Array-of-Structures (AoS) and 
Structure-of-Arrays (SoA), are also used to represent triangular meshes in 
our implementation. Furthermore, the feature CUDA Dynamic Parallelism (CDP) 
is employed to realize the nested parallelization in Smart Laplacian 
Smoothing. The feature CDP is an extension to the CUDA programming model which enables a 
CUDA kernel to create and synchronize with new kernel(s) directly on the GPU \cite{24nvidaia}. We finally carry out several experimental tests to evaluate the 
performance of our GPU-accelerated implementation by comparing to the 
corresponding CPU implementation.


\section{Background}
\label{sec:background}

\subsection{Laplacian Smoothing}

\begin{figure}[!h]
	\centering
	\includegraphics[scale = 0.85]{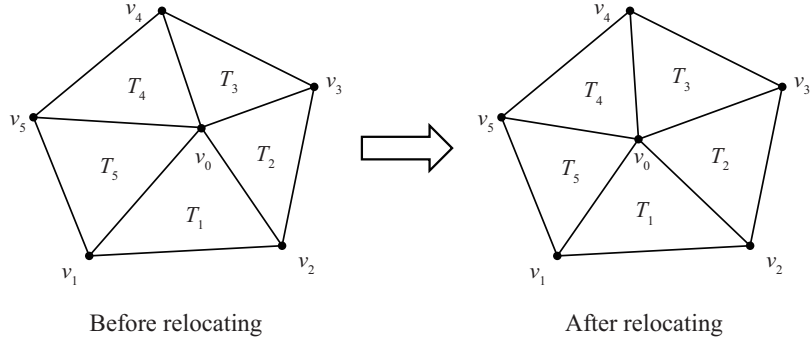}
	\caption{An illustration of Laplacian smoothing}
	\label{fig:fig1}
\end{figure}

Laplacian smoothing is one of the most commonly used mesh smoothing 
algorithms \cite{15herman1976}. The basic idea behind Laplacian smoothing is to relocate 
each node in a mesh to the geometric center of its neighboring nodes; see 
Figure \ref{fig:fig1}.

Laplacian smoothing is computationally straightforward but does not always 
produces enhancements in mesh quality. In practical applications, inverted or 
even invalid elements in concave regions are probably created. To deal with 
the above problem, some variations such as the \textit{Weighted} Laplacian smoothing \cite{16blacker1991,17vollmer} and \textit{Constrained} / \textit{Smart} Laplacian smoothing \cite{9canann,18wei} have 
been designed.

The basic idea behind Smart Laplacian smoothing is also simple. For a node 
being smoothed such as the one $v_0 $ presented in Figure \ref{fig:fig1}, the mesh 
quality of its incident elements (those elements that share this node, e.g., 
the triangles $T_1 $, $T_2 $, $T_3 $, $T_4 $, and $T_5 $ in Figure \ref{fig:fig1}) is 
first evaluated; Then a newly smoothed position of this node is calculated 
according to the Laplacian smoothing operations. The quality of all the 
incident elements is evaluated again using the new nodal position. If the 
mesh quality is increased, then the new nodal position will be accepted; 
otherwise, the node will not be repositioned.

\subsection{Iteration Forms}

When calculating the smoothed coordinates of vertices according to the 
smoothing operation, there are typically two forms in terms of selecting the 
coordinates of neighboring nodes \cite{1mei2014}. These two forms can be simply 
illustrated using the following formulations.

Form A :
\begin{equation}
\label{eq2}
\overline {x_i^{q+1} } =\frac{1}{N}\sum\limits_{j=1}^N {x_j^q }, 
\end{equation}
where~$N$~is the number of neighboring nodes to node~$i$~and~$\overline 
{x_i^{q+1} } $~is the new position for node~$i$ in the iteration pass ($q$ + 1).

Form B:
\begin{equation}
\label{eq3}
\overline {x_i^{q+1} } =\frac{1}{N}\left( {\sum\limits_{j=1}^{N_q } {x_j^q } 
	+\sum\limits_{k=1}^{N_{q+1} } {x_k^{q+1} } } \right),\mbox{ }\left\{ 
{{\begin{array}{*{20}c}
		{\mbox{0}\le N_q \le N} \\
		{\mbox{0}\le N_{q+1} \le N} \\
		{N_q +N_{q+1} =N} \\
		\end{array} }} \right.,
\end{equation}
where~$N$~is the number of neighboring nodes to node~$i$~and~$\overline 
{x_i^{q+1} } $~is the new position for node~$i$ in the iteration pass ($q$ + 1). 
$N_q$ and $N_{q+1}$ are numbers of neighboring nodes derived from the 
iteration passes $q$ and ($q$+1), respectively. Obviously, the Form A is a special 
case of the Form B where $N_{q+1} =0$.

\subsection{Data Layouts}

\begin{figure}[h!]
	\centering
	\subfigure[AoS]{
		\label{fig:fig2:a}       
		\includegraphics[scale = 0.8]{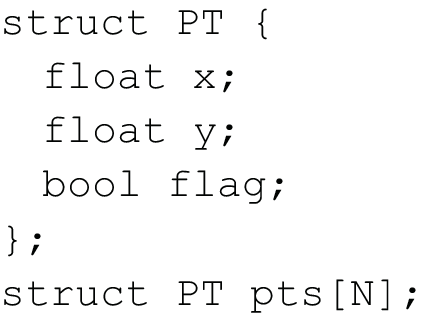}
	}
	\hspace{3em}
	\subfigure[SoA]{
		\label{fig:fig2:b}       
		\includegraphics[scale = 0.8]{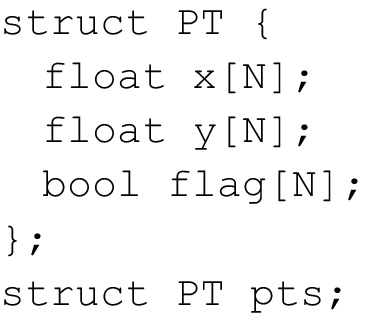}
	}
	\caption{The data layouts: Array-of-Structures (AoS) and Structure-of-Arrays (SoA)}
	\label{fig:fig2}       
\end{figure}

In GPU-accelerated applications there are typically two commonly used data 
layouts, including the Array-of-Structures (AoS) and Structure-of-Arrays 
(SoA); see a simple illustration in Figure \ref{fig:fig2}. Arranging data in AoS layout 
leads to coalescing issues as the data are interleaved. Arranging data as 
the SoA layout makes full use of the memory bandwidth even when individual 
elements of the structure are utilized. In practical applications, it is unable to determine which data layout can 
always achieve better efficiency. In general, the selecting of the proper 
data layout is application-specific. 

\section{Our Implementation}
\label{sec:implement}

\subsection{Overview}

Our implementation of the Smart Laplacian smoothing is mainly divided 
into four sub-procedures, including the initiating procedure, the finding of 
neighbors, the determining of constraints, and the iterative procedure to 
obtain the smoothed positions. The initiating 
procedure is to set several values for each vertex and calculate the 
mesh quality for each triangle. The finding of 
neighbors is to identify the neighboring vertices / nodes
and to find the incident elements / triangles for each vertex. The determining of constraints is to identify 
which vertices are needed to be fixed or free. Finally, the iterating procedure is to 
iteratively calculate the positions of smoothed points. 

\subsection{Data Structures}

We design two sets of mesh data structures according to the two layouts AoS 
and SoA, respectively. Corresponding 
implementations are also developed 
according to these mesh data structures. Due to the limit of the paper length, we only list the data structures represented with the SoA layout; see Listing \ref{code:soa}.

\begin{lstlisting}[
frame = tb,
caption =  Mesh data structures represented with the SoA layout,
label = code:soa
%label = code:soa,
%float
]
struct cuVert_SOA_ST {
  float *x, *y;      // Coordinates of a node
  int *nNeig, *neig; // Number, and indices of neighboring nodes
  int *nLoca, *loca; // Number, and indices of incident elements
  bool *bBoundary;   // Whether nodes are boundary vertices
  float *minQuality; // Qalities of the worst incident elements
};
struct cuTrgl_SOA_ST {
  int *vIDs;         // Indices of vertices of triangles
  float *Quality;    // Qualities of triangles
};
struct cuMesh_SOA_ST {
  int nVert, nTrgl;  // Number of vertices and triangles
  cuVert_SOA_ST verts;  cuTrgl_SOA_ST trgls;
};
\end{lstlisting}

\subsection{Implementation Details}

\subsubsection{The Initiating Procedure}

This procedure is to (1) calculate the mesh quality metric of each triangle 
for the first time, and (2) initiate several values to prepare for the 
subsequent calculations. We adopt the mesh quality metric, 
\textit{$\alpha $, }proposed by Lee and Lo \cite{25lee} to measure the quality of triangular elements. 

We design a simple CUDA kernel to calculate the qualities of all triangular 
elements, where each thread is responsible for computing the quality of a 
single triangle. The element quality is stored in the component
\texttt{float *Quality} in the structure 
\texttt{cuTrgl{\_}SOA{\_}ST}.

We also design another simple kernel to initiate the values of the 
components \texttt{nNeig}, \texttt{nLoca}, and \texttt{bBoundary} as 0, 0, and \texttt{false}, respectively. 
Each thread within this kernal / grid is responsible for initiating the 
values for only one points. The above set values mean that:
currently there are no found adjacent points and incident triangles for any vertex. In 
addition, any vertex of the mesh is initially set to be free point that can 
be relocated in smoothing. 

\subsubsection{The Finding of Neighbors}

The finding of neighbors is to find: (1) the adjacent / neighboring nodes 
and (2) the incident triangles for each vertex. As described in our previous 
work \cite{1mei2014}, the finding of neighbors can be quite easily performed 
according to the topology of the mesh. More specifically, for a triangle the 
second and the third vertices must be the adjacent points of the first 
vertex; similarly, the first and the third vertices are definitely the 
adjacent points of the second vertex. And obviously, each triangle is the 
incident triangle of its three vertices. 

As also explained in our previous work \cite{1mei2014}, due to the data 
dependencies, we allocate only one thread block and a single thread within 
the thread block to perform the finding of neighbors. The only one thread 
takes the complete responsibilities to finding the adjacent points and 
incident triangles for all vertices. In fact, this procedure carried out on 
the GPU is the same as the corresponding version performed on the CPU. 

\subsubsection{The Determining of Constraints}

The determining of constraints is to identify which vertices of a mesh can 
be relocated in smoothing and which cannot be. For planar polygonal meshes, the 
constraints include the boundary vertices and other specifically-defined 
points such as some feature points. In our implementation, we only consider 
the boundary vertices as constraints. 

We also adopt the method introduced in \cite{1mei2014} to 
determine the boundary vertices. More specifically, by taking advantage of 
the indices of the adjacent points, it is quite easy to check whether or not 
a vertex is a boundary one. The basic idea behind determining the boundary 
vertices is straightforward: if all the neighbors of a vertex, e.g., the 
node $v_0$ in Figure \ref{fig:fig1}, have been recorded twice, then the vertex is internal; 
otherwise, it is a boundary vertex.

We develop a specific kernel to carry out this determine of boundary 
vertices. Each thread within the thread grid is invoked to check whether a points is 
a boundary one, i.e., to check whether all the neighbors / adjacent points 
of a vertex has been recorded twice. If not, then update the corresponding 
flag value \texttt{bool bBoundary} from being \texttt{false} to \texttt{true}. 

\subsubsection{The Iterating Procedure}

The final and key procedure is to iteratively calculate the smoothed 
positions of all vertices. The main feature of this procedure is that: only 
one thread block is allocated because of the data dependencies. In Smart 
Laplacian smoothing, the calculating of the positions of all smoothed points 
in the ($i+1$)$^{th}$ iteration depends on the positions of all smoothed points 
in the $i^{th}$ iteration. In other words, there exist data dependencies 
between the ($i+1$)$^{th}$ iteration and the $i^{th}$ iteration. 

Due to the data dependencies existing in different passes of iterations, 
only one thread block is allocated. Each thread within the thread block is 
invoked to calculate the smoothed positions of \texttt{(n + BLOCK{\_}SIZE - 1) / 
BLOCK{\_}SIZE} vertices in one iterative step, where \texttt{n} is the number of all 
vertices and \texttt{BLOCK{\_}SIZE} denotes the number of threads within the only one 
thread block. The barrier of synchronization \texttt{{\_}{\_}syncthreads()} is used 
to guarantee all threads within the only one block finishing calculating one 
pass of all smoothed positions.

In Smart Laplacian smoothing it is ``Smart'' 
to determine whether or not a non-constrained vertex should be relocated. 
This ``Smart'' determination is typically carried out by comparing the mesh 
quality of the original mesh before relocating the vertex and the mesh 
quality after the relocating. In other words, when using the new nodal 
coordinates after relocating, if the mesh quality of the local mesh (i.e., 
all the incident triangles) of the vertex being smoothed definitely 
improves, for example, if the min value of the mesh quality metric of all 
the incident triangles is increased, then this vertex should be 
repositioned; otherwise, the position of the vertex will be leaved unchanged. 
Therefore, after obtaining the new position of a smoothed vertex, it is 
needed to temporarily re-evaluate the mesh quality of the local mesh, and 
then compare the mesh qualities. 

After completely obtaining all the smoothed positions in one iteration step, 
the mesh quality of all triangles is needed to be updated using all of the 
new nodal coordinates. The qualities of all triangles updated in $i^{th}$ 
iteration will be used to ``Smart'' calculate the new smoothed positions in 
the ($i+1$)$^{th}$ iteration. 
We implement the above updating of mesh quality in each iteration step by 
with or without the use of the feature, CUDA Dynamic Parallelism (CDP).

\textit{``GPU-'' versions}: In this version, the feature CDP is not 
adopted. Alternatively, in this kernel of iterating, each thread within the 
only thread block is responsible for: (1) calculating the quality of all of 
its incident triangles, and (2) finding the min value of the qualities of 
the incident triangles. The min quality is then stored in the component 
\texttt{float minQuality} of the data structure 
\texttt{cuVert{\_}SOA{\_}ST}. 

\textit{``CDP-'' versions}: In this version, the feature CDP is used. We 
design a child kernel specifically for updating the quality of all triangles 
 by using new nodal coordinates. Within this child kernel, each 
thread is invoked to evaluate the quality of only one triangle. We also 
develop another child kernel for finding the min quality of the incident 
triangles for all vertices. Each thread in this kernel is responsible for 
finding the min values of the quality of the incident triangles. Note that the above two 
child kernels are only invoked once in the parent kernel of iterating.


\section{Results}
\label{sec:results}

To evaluate the performance of our GPU implementations, we perform the 
experimental tests on the GeForce GT640 (GDDR5) graphics cards with CUDA 6.5. The CPU experiments are performed on Windows 7 SP1 with a dual Intel i5 3.2 GHz processor and 8GB of RAM memory.We perform the 
experimental tests for both the GPU- and CDP- versions of our 
implementation. 
Each version of our implementation is tested in two cases where the 
iteration Form A and Form B are adopted. 

The test data includes 5 planar triangular meshes  which are composed of 1, 5, 10, 50, and 100K vertices, respectively. The original unsmoothed 
triangular meshes are generated according to the standard Delaunay 
triangulation algorithm. First, five sets of uniformly distributed points in 
2D are randomly generated using the generator provided by Qi, et al. 
\cite{26qi}; and then Delaunay meshes are created for these sets of discrete points 
using the library Triangle \cite{27shewchuk}. 

We evaluate our implementation on the single precision. For the GPU-version, 
the running time and corresponding speedups achieved when using the Form A 
and Form B are listed in Table \ref{tab:tab6} and Table \ref{tab:tab7}, respectively. Similarly, For the CDP-version, the running time and corresponding 
speedups obtained when iterating in the Form A and Form B are presented in 
Table \ref{tab:tab8} and Table \ref{tab:tab9}, respectively. 
We find that the highest speedup is up to 44.76x; see Table \ref{tab:tab8}.

\begin{table}[!h]
	\caption{Performance of the GPU-version developed in the iteration Form A}
	\begin{center}
        \begin{tabular}{p{50pt}p{50pt}p{50pt}p{50pt}p{5pt}p{50pt}p{50pt}}
			\toprule
			\raisebox{-1.50ex}[0cm][0cm]{Size}& 
			\multicolumn{3}{c}{Running time (/ms)} & 
			\multicolumn{3}{c}{Speedup}  \\
			\cmidrule(r){2-4} 
			\cmidrule(r){6-7} 
			& 
			CPU& GPU-AoS& GPU-SoA&& GPU-AoS& GPU-SoA \\
			\midrule
			1K& 218& 16.0& 18.8&& 13.63& 11.60 \\
			5K& 1217& 72.2& 86.0&& 16.86& 14.15 \\
			10K& 2012& 117.6& 138.1&& 17.11& 14.57 \\
			50K& 14383& 530.5& 618.0&& 27.11& 23.27 \\
			100K& 33836& 802.8& 932.8&& 42.15& 36.27 \\
			\bottomrule
		\end{tabular}
		\label{tab:tab6}
	\end{center}
\end{table}

\begin{table}[!h]
	\caption{Performance of the GPU-version developed in the 
		iteration Form B}
	\begin{center}
        \begin{tabular}{p{50pt}p{50pt}p{50pt}p{50pt}p{5pt}p{50pt}p{50pt}}
			\toprule
			\raisebox{-1.50ex}[0cm][0cm]{Size}& 
			\multicolumn{3}{c}{Running time (/ms)} & 
			\multicolumn{3}{c}{Speedup}  \\
			\cmidrule(r){2-4} 
			\cmidrule(r){6-7} 
			& 
			CPU& GPU-AoS& GPU-SoA&& GPU-AoS& GPU-SoA \\
			\midrule
			1K& 171& 14.9& 17.7&& 11.48& 9.66 \\
			5K& 1139& 52.4& 76.6&& 21.74& 14.87 \\
			10K& 1888& 79.3& 115.0&& 23.81& 16.42 \\
			50K& 11918& 424.5& 513.1&& 28.08& 23.23 \\
			100K& 26021& 776.1& 896.3&& 33.53& 29.03 \\
			\bottomrule
		\end{tabular}
		\label{tab:tab7}
	\end{center}
\end{table}

\begin{table}[!h]
	\caption{Performance of the CDP-version developed in the iteration Form A}
	\begin{center}
        \begin{tabular}{p{50pt}p{50pt}p{50pt}p{50pt}p{5pt}p{50pt}p{50pt}}
			\toprule
			\raisebox{-1.50ex}[0cm][0cm]{Size}& 
			\multicolumn{3}{c}{Running time (/ms)} & 
			\multicolumn{3}{c}{Speedup}  \\
			\cmidrule(r){2-4} 
			\cmidrule(r){6-7} 
			& 
			CPU& GPU-AoS& GPU-SoA&& GPU-AoS& GPU-SoA \\
			\midrule
			1K& 218& 14.1& 16.5&& 15.46& 13.21 \\
			5K& 1217& 59.2& 66.8&& 20.56& 18.22 \\
			10K& 2012& 96.1& 107.7&& 20.94& 18.68 \\
			50K& 14383& 477.7& 532.4&& 30.11& 27.02 \\
			100K& 33836& 755.9& 832.5&& 44.76& 40.64 \\
			\bottomrule
		\end{tabular}
		\label{tab:tab8}
	\end{center}
\end{table}

\begin{table}[!h]
	\caption{Performance of the CDP-version developed in the iteration Form B}
	\begin{center}
        \begin{tabular}{p{50pt}p{50pt}p{50pt}p{50pt}p{5pt}p{50pt}p{50pt}}
			\toprule
			\raisebox{-1.50ex}[0cm][0cm]{Size}& 
			\multicolumn{3}{c}{Running time (/ms)} & 
			\multicolumn{3}{c}{Speedup}  \\
			\cmidrule(r){2-4} 
			\cmidrule(r){6-7} 
			& 
			CPU& GPU-AoS& GPU-SoA&& GPU-AoS& GPU-SoA \\
			\midrule
			1K& 
			171& 
			13.5& 
			15.4&& 
			12.67& 
			11.10 \\
			5K& 
			1139& 
			50.0& 
			58.1&& 
			22.78& 
			19.60 \\
			10K& 
			1888& 
			74.8& 
			82.8&& 
			25.24& 
			22.80 \\
			50K& 
			11918& 
			404.4& 
			424.8&& 
			29.47& 
			28.06 \\
			100K& 
			26021& 
			750.2& 
			801.7&& 
			34.69& 
			32.46 \\
			\bottomrule
		\end{tabular}
		\label{tab:tab9}
	\end{center}
\end{table}


\section{Discussion}
\label{sec:discuss}

We have investigated the related work involving the accelerating of mesh 
smoothing on the GPU, and have found that currently there are only several 
recent efforts \cite{1mei2014,5damato2013,6dahal}. Among the above efforts, the work presented in \cite{5damato2013} focused on smoothing tetrahedral meshes, while in \cite{1mei2014} 
and \cite{6dahal} the work aimed at developing the GPU-accelerated mesh 
smoothing for planar meshes.

The GPU-accelerated implementations of 2D Laplacian mesh 
smoothing were both introduced in \cite{1mei2014} and \cite{6dahal}. However, the 
above implementations are only developed for the original Laplacian 
smoothing, rather than the Smart Laplacian smoothing. To the best of the 
authors' knowledge, the work presented in this paper is the first attempt at 
accelerating Smart Laplacian smoothing on the GPU.

\subsection{Impact of Data Layouts (AoS and SoA)}

We have observed that: the version of our implementation that is 
developed using the data layout AoS is slightly faster than that is 
developed using the layout SoA. This behavior has previously been
observed in our previous work \cite{1mei2014}. As also explained in our previous work, 
the better performance obtained by the GPU version based upon the AoS data 
structures is due to the use of the aligned global memory accesses. Therefore, in 
practical applications, we recommend the developers to use the data layout 
AoS to implement the Smart Laplacian smoothing. However, it is should be also noted that: the design of the AoS format mesh data structures is much more complex than that of the SoA formant.

\subsection{Impact of Iteration Forms}

\begin{figure}[h!]
	\centering
	\subfigure[AoS]{
		\label{fig:fig7:a}       
		\includegraphics[scale = 0.45]{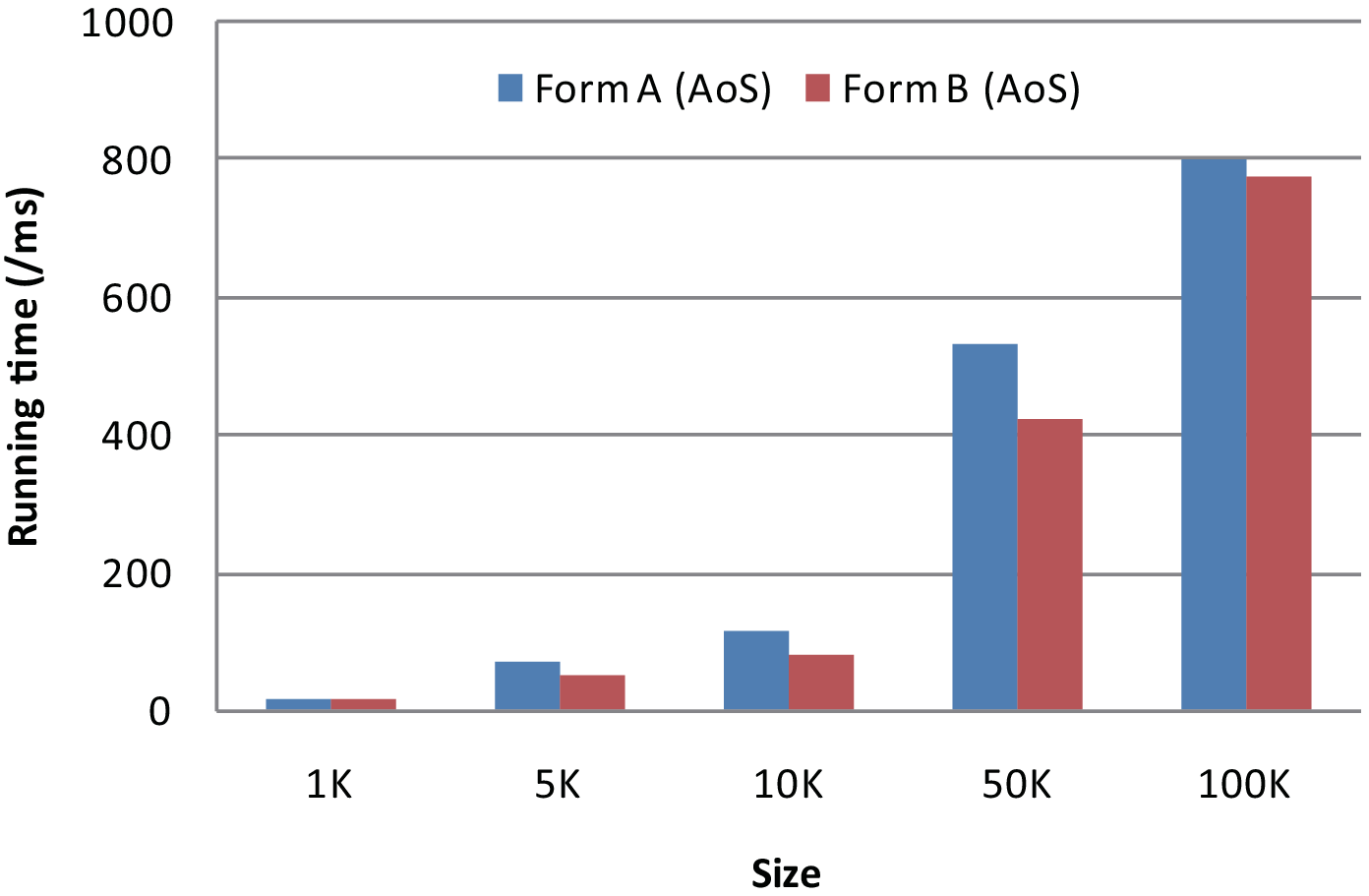}
	}
	\hspace{1em}
	\subfigure[SoA]{
		\label{fig:fig7:b}       
		\includegraphics[scale = 0.45]{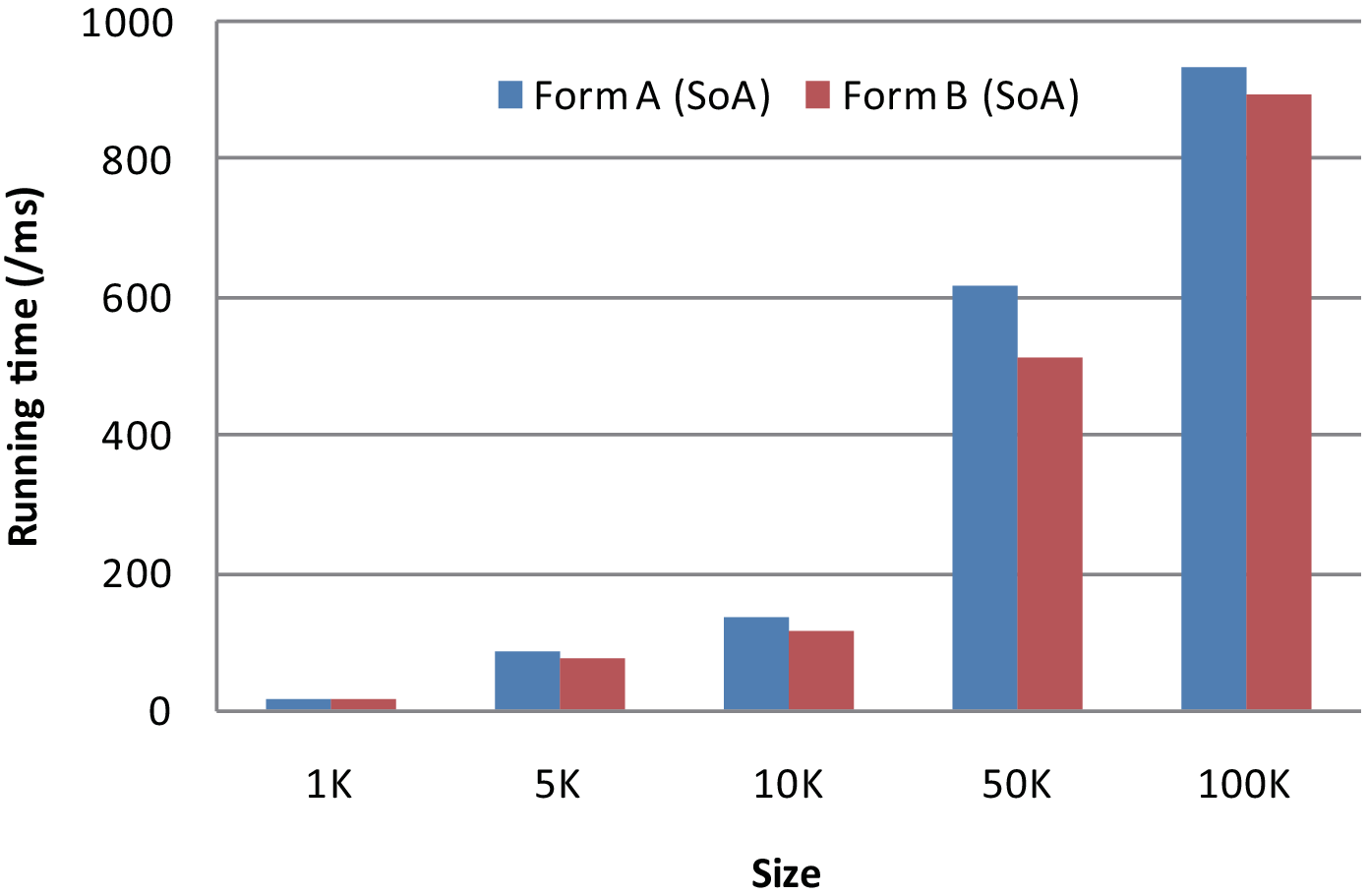}
	}
	\caption{Comparison of the running time of two variations implemented in 
		the Form A and Form B}
	\label{fig:fig7}       
\end{figure}

By comparing the absolute running time of two variations developed using the 
two iteration forms, i.e., the Form A and the Form B, we have found that: 
the variation with the Form B is a little faster than the variation with the 
Form A in the cases whenever which data layour (AoS or SoA) is adopted or 
whether the feature CDP is used; see Figure \ref{fig:fig7} for a groups of performance 
comparison. 

The causes to the above results have been also described in our previous 
work \cite{1mei2014}. The first cause is that: the Form A needs more 
calculations due to swapping intermediate nodal coordinates during 
iterating. The other cause is that: the convergence speed of the Form A is 
much lower than that of the Form B; and thus the Form A needs much more 
iterations for converging. Therefore, the Form B is suggested in practical 
applications. 

\subsection{Performance of the Use of CUDA Dynamic Parallelism}

We have observed that: the ``CDP'' version is slightly faster than the 
``GPU'' version. This is perhaps leaded by the following causes. In the GPU 
version, only a thread block is allocated; and after obtaining the new 
positions of interior points, the quality of local mesh for each point is 
needed to be re-evaluated. Within this only one thread block, a single 
thread takes the responsibilities for re-calculating the quality of local 
meshes for several points, rather than only one point. In this case, the 
power of massively parallel computing is not fully exploited. The second 
reason is that: when re-evaluating the quality of the local mesh for a 
single point, each of the incident elements is needed to evaluate it quality once. Due to the fact that a 
triangle has three vertices, it is thus needed to evaluate its quality three 
times. Obviously, there exists redundancies in the calculating of the quality of 
triangles.

\begin{figure}[h!]
	\centering
	\subfigure[Using the layout SoA in Form A]{
		\label{fig:fig8:soa:FormA}       
		\includegraphics[scale = 0.45]{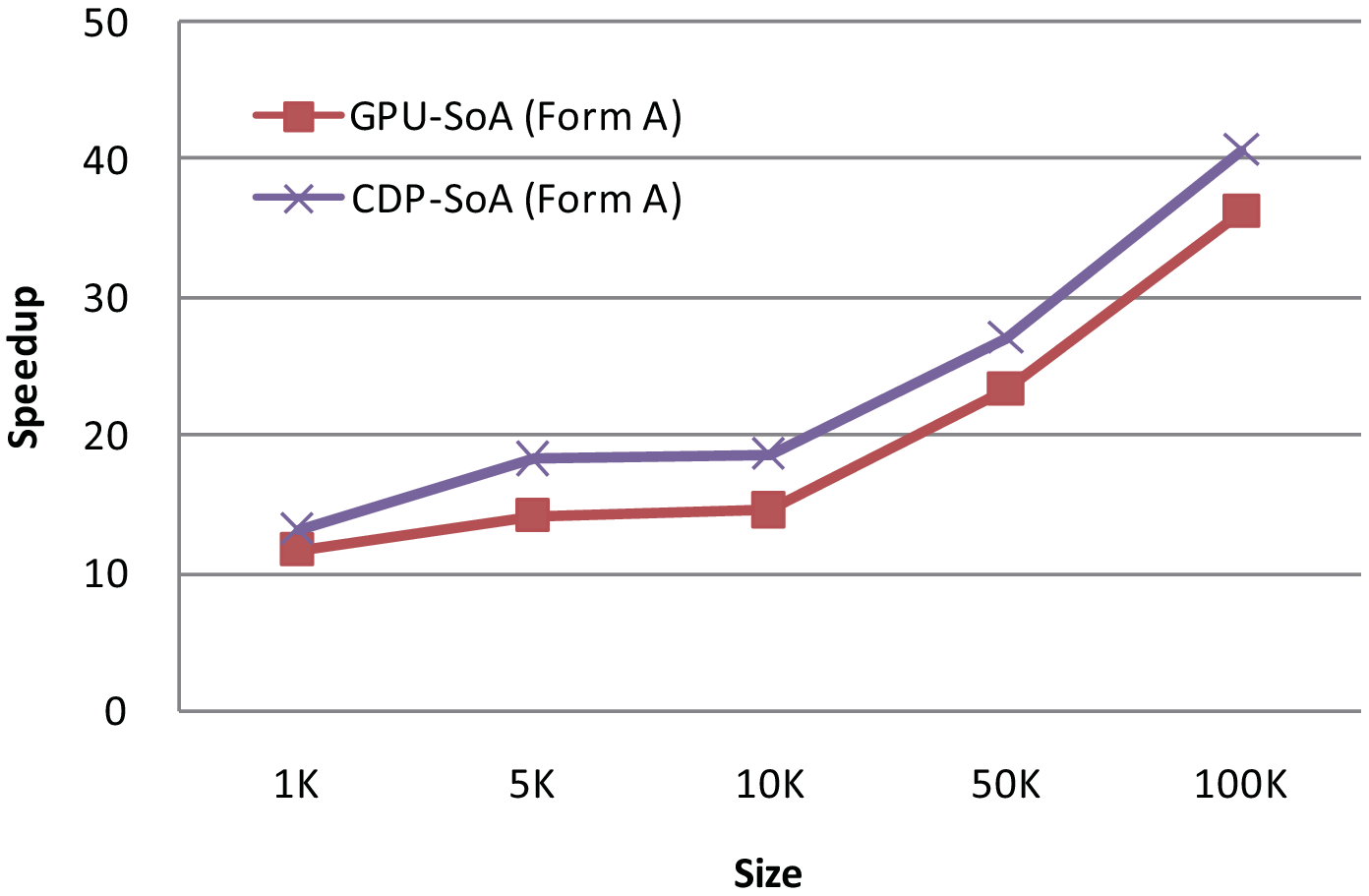}
	}
	\hspace{1em}
	\subfigure[Using the layout AoS in Form A]{
		\label{fig:fig8:aos:FormA}       
		\includegraphics[scale = 0.45]{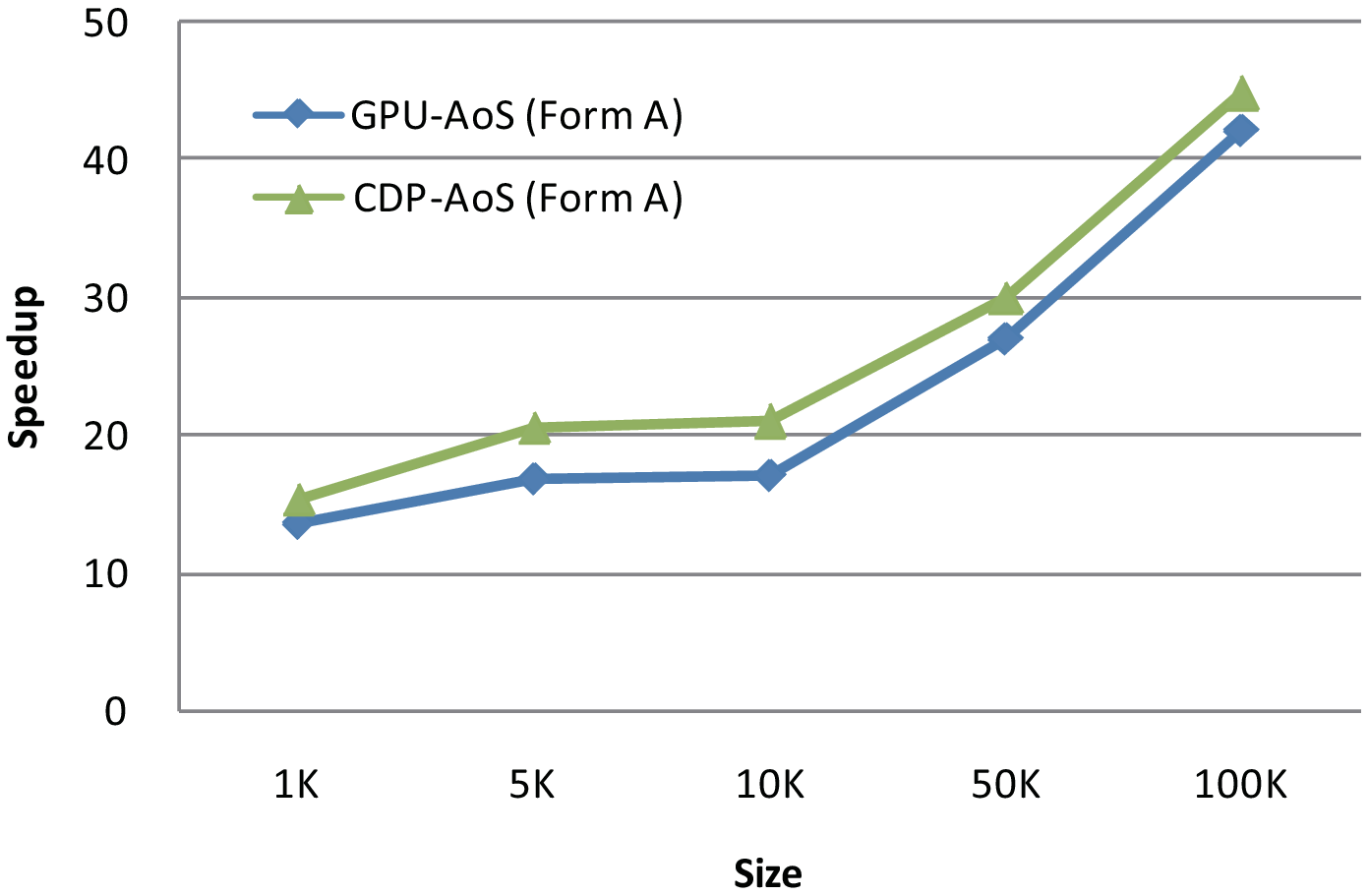}
	}
	\caption{Comparison of the speedups of the GPU-version and CDP-version of 
		our implementation}
	\label{fig:fig8}       
\end{figure}

In the CDP version, we first design a specific kernel to re-calculate the 
quality of all triangles after obtaining the new positions in each 
iteration. Each thread within this kernel is responsible for computing the 
quality metric of only one triangle. We also develop another kernel to find 
the worst triangle with the lowest quality of the local mesh for each point. For that the quality of 
each of the incident elements (i.e., triangles) have been calculated, it is 
quite easy to find the worst triangle that has the lowest quality among the 
incident elements. In this kernel, each thread is designed to find the worst 
incident element for only one point. Obviously, in this CDP version, the 
power of the massively parallel computing in the above two child kernels / 
grids can be efficiently exploited. However, the computing capability of the 
parent kernel is not fully exploited since only one thread (typically the first 
thread within the thread grid) is needed to invoke the above two child 
kernels. This is probably the reason why the CDP version is slightly 
rather than significantly faster than the GPU version.


\section{Conclusion}
\label{sec:conclusion}

We have presented an efficient GPU-accelerated implementation of the Smart 
Laplacian smoothing for optimizing planar triangular meshes. We have 
developed our implementation by using two data layouts, Array-of-Structures 
(AoS) and Structure-of-Arrays (SoA), and employing two iteration forms (Form 
A and Form B). The feature CDP is also adopted to realize the nested 
parallelization to iteratively determine the smoothed vertices' positions. 
We have evaluated the performance of our implementation using five randomly 
created triangular meshes on the GPU GT640. Experimental results 
have indicated that: our implementation can achieve the speedups of up to 44x. We have also found 
that: the data layout AoS can obtain better efficiency than the SoA layout. 
It has been demonstrated that: the Form A that needs to swap intermediate 
nodal coordinates is always slower than the Form B that does not 
swap data.We have also observed that: the version of our 
implementation with the use of the feature CDP is slightly faster than 
the version where the CDP is not adopted. 



\vspace*{10pt}
\textbf{Acknowledgments}
The authors are grateful to the anonymous referee 
for helpful comments that improved this paper. This research was supported 
by the Natural Science Foundation of China (Grant Nos. 40602037 and 
40872183).

%
\label{sect:bib}
\bibliographystyle{plain}
\bibliography{myBib}

\begin{thebibliography}{10}

\bibitem{13beniterz}
Domingo Benitez, Eduardo Rodríguez, JoséM. Escobar, and Rafael Montenegro.
\newblock The effect of parallelization on a tetrahedral mesh optimization
  method.
\newblock In {\em Parallel Processing and Applied Mathematics}, Lecture Notes
  in Computer Science, pages 163--173. Springer Berlin Heidelberg, 2014.

\bibitem{14benitez}
Domingo Benítez, Eduardo Rodríguez, JoséMaría Escobar, and Rafael
  Montenegro.
\newblock Performance evaluation of a parallel algorithm for simultaneous
  untangling and smoothing of tetrahedral meshes.
\newblock In {\em Proceedings of the 22nd International Meshing Roundtable},
  pages 579--598. Springer International Publishing, 2014.

\bibitem{16blacker1991}
Ted~D. Blacker and Michael~B. Stephenson.
\newblock {Paving: A new approach to automated quadrilateral mesh generation}.
\newblock {\em International Journal for Numerical Methods in Engineering},
  32(4):811--847, 1991.

\bibitem{9canann}
Scott~A Canann, Joseph~R Tristano, and Matthew~L Staten.
\newblock {An approach to combined Laplacian and optimization-based smoothing
  for triangular, quadrilateral, and quad-dominant meshes}.
\newblock In {\em Proceedings of the 7th International Meshing Roundtable},
  pages 479--494. 1998.

\bibitem{4chen}
Junjie Chen, Xiaogang Jin, and Zhigang Deng.
\newblock {GPU-based polygonization and optimization for implicit surfaces}.
\newblock {\em The Visual Computer}, pages 1--12, 2014.

\bibitem{6dahal}
Sangeet Dahal and Timothy~S. Newman.
\newblock {Efficient, GPU-based 2D mesh smoothing}.
\newblock In {\em Proceedings of the IEEE SOUTHEASTCON 2014}, pages 1--7. 2014.

\bibitem{5damato2013}
J.~P. D'Amato and M.~Venere.
\newblock {A CPU-GPU framework for optimizing the quality of large meshes}.
\newblock {\em Journal of Parallel and Distributed Computing},
  73(8):1127--1134, 2013.

\bibitem{3damato}
JP~D'amato and P~Lotito.
\newblock {Mesh optimization with volume preservation using GPU}.
\newblock {\em Latin American applied research}, 41(3):291, 2011.

\bibitem{11field1988}
David~A. Field.
\newblock {Laplacian smoothing and Delaunay triangulations}.
\newblock {\em Communications in Applied Numerical Methods}, 4(6):709--712,
  1988.

\bibitem{Gorman2012}
G.J. Gorman, J.~Southern, P.E. Farrell, M.D. Piggott, G.~Rokos, and P.H.J.
  Kelly.
\newblock {Hybrid OpenMP/MPI anisotropic mesh smoothing}.
\newblock {\em Procedia Computer Science}, 9:1513 -- 1522, 2012.
\newblock Proceedings of ICCS2012.

\bibitem{15herman1976}
Leonard~R Herrmann.
\newblock {Laplacian-isoparametric grid generation scheme}.
\newblock {\em Journal of the Engineering Mechanics Division}, 102(5):749--907,
  1976.

\bibitem{25lee}
CK~Lee and SH~Lo.
\newblock {A new scheme for the generation of a graded quadrilateral mesh}.
\newblock {\em Computers \& structures}, 52(5):847--857, 1994.

\bibitem{1mei2014}
Gang Mei, John~C. Tipper, and Nengxiong Xu.
\newblock {A generic paradigm for accelerating Laplacian-based mesh smoothing
  on the GPU}.
\newblock {\em Arabian Journal for Science and Engineering}, 39(11):7907--7921,
  2014.

\bibitem{24nvidaia}
NVIDIA.
\newblock {CUDA C Programming Guide v6.5}, 2014.

\bibitem{26qi}
Meng Qi, Thanh-Tung Cao, and Tiow-Seng Tan.
\newblock {Computing 2D constrained Delaunay triangulation using the GPU}.
\newblock {\em Visualization and Computer Graphics, IEEE Transactions on},
  19(5):736--748, 2013.

\bibitem{27shewchuk}
Jonathan~Richard Shewchuk.
\newblock {Triangle: Engineering a 2D quality mesh generator and Delaunay
  triangulator}.
\newblock In {\em Applied Computational Geometry Towards Geometric
  Engineering}, pages 203--222. Springer Berlin Heidelberg, 1996.

\bibitem{10vartz}
Dimitris Vartziotis and Benjamin Himpel.
\newblock {Efficient and global optimization-based smoothing methods for
  mixed-volume meshes}.
\newblock In Josep Sarrate and Matthew Staten, editors, {\em Proceedings of the
  22nd International Meshing Roundtable}, pages 293--311. Springer
  International Publishing, 2014.

\bibitem{7vartz}
Dimitris Vartziotis and Benjamin Himpel.
\newblock {Efficient mesh optimization using the gradient flow of the mean
  volume}.
\newblock {\em SIAM Journal on Numerical Analysis}, 52(2):1050--1075, 2014.

\bibitem{17vollmer}
J.~Vollmer, R.~Mencl, and H.~Muller.
\newblock {Improved Laplacian smoothing of noisy surface meshes}.
\newblock {\em Computer Graphics Forum}, 18(3):131--138, 1999.

\bibitem{18wei}
Mingqiang Wei, Wuyao Shen, Jing Qin, Jianhuang Wu, Tien-Tsin Wong, and
  Pheng-Ann Heng.
\newblock {Feature-preserving optimization for noisy mesh using joint bilateral
  filter and constrained Laplacian smoothing}.
\newblock {\em Optics and Lasers in Engineering}, 51(11):1223--1234, 2013.

\bibitem{2zegard}
Tomás Zegard and GlaucioH Paulino.
\newblock {Toward GPU accelerated topology optimization on unstructured
  meshes}.
\newblock {\em Structural and Multidisciplinary Optimization}, 48(3):473--485,
  2013.

\end{thebibliography}


\end{document}